\documentclass[amsmath,twocolumn,prl,aps]{revtex4-1}
\usepackage{amsthm,amsfonts,graphicx,verbatim, color}
\usepackage{bm}
\usepackage[utf8]{inputenc}

\newcommand{\alphaCF}{\ensuremath{\alpha_{CF}}}
\newcommand{\alphaF}{\ensuremath{\alpha_{F}}}
\newcommand{\vq}{\ensuremath{\vec q}}

\begin{document}

\title{Numerical Study of Anisotropy in the Composite Fermi liquid}

\author{Matteo Ippoliti, Scott D. Geraedts, and R. N. Bhatt}
\affiliation{Departments of Electrical Engineering and Physics, Princeton University, Princeton NJ 08544, USA}
\begin{abstract}
We perform density-matrix renormalization group studies of a two-dimensional electron gas in a high magnetic field and with an anisotropic band mass.
At half-filling in the lowest Landau level, such a system is a Fermi liquid of composite fermions.
By measuring the Fermi surface of these composite fermions, we determine a relationship between the anisotropy of composite fermion dispersion, $\alphaCF$, and the original anisotropy $\alphaF$ of the fermion dispersion at zero magnetic field. For systems where the electrons interact via a Coulomb interaction, we find $\alphaCF=\sqrt{\alphaF}$ within our numerical accuracy. The same result has been found concurrently in recent experiments.
We also find that the relationship between the anisotropies is dependent on the form of the electron-electron interaction.
\end{abstract}
\maketitle

Two dimensional electron systems at the interface of semiconductors in strong magnetic fields host a wide variety of exotic Abelian and non-Abelian gapped fractional quantum Hall phases~\cite{GirvinReview,SternReview,JainBook}, 
as well as broken symmetry phases such as the nematic, Wigner crystal and bubble phases~\cite{FradkinReview,Fogler2002,Mulligan2010,Xia2011,Koduvayur2011,You2014,Samkharadze2016,You2016}.
These systems have received a lot of interest in the condensed matter physics community, both because of the elegant mathematical structure emanating from Landau level physics, and also because of their ready accessibility under multiple experimental platforms. 
Unique among the quantum Hall phases is the gapless phase at half filling in the lowest Landau level, identified as the `composite Fermi liquid' (CFL)~\cite{HLR}.
Recently, there has been a revival of interest~\cite{Son2015,Geraedts} in the CFL, which has a Fermi surface, analogous to the Fermi liquid phase at zero field.
\footnote{Much of this interest has to do with the nature of particle-hole symmetry in the CFL, which is not the subject of this paper. To our knowledge the recent `Dirac composite fermion picture'\cite{Son2015} should not affect the response to anisotropy measured here.}

Much of the theoretical work on this subject has assumed rotational symmetry, though this symmetry is not central to the physics of quantum Hall systems.
Understanding the quantum Hall effect in the absence of rotational symmetry is an active area of research~\cite{Joynt,Haldane2011,Qiu2012,Papic2013,Johri2016,Haldane2016}.
Experimentalists have been able to break rotational symmetry in several ways, {\it e.g.}~by applying parallel magnetic fields~\cite{Lilly1999,Liu2013,Kamburov2013,*Kamburov2014,*Mueed2015,*Mueed2016}, 
straining their samples~\cite{Jo2017}, 
or doing experiments on materials which have anisotropic Fermi surfaces in the absence of magnetic fields, such as many-valley semiconductors~\cite{Shayegan2006,Gokmen2010}.
It is therefore very interesting to ask what happens to the various well-understood states of matter when rotational symmetry is broken. In this work, we investigate the effect of breaking rotational symmetry on the gapless composite Fermi liquid state for a half-filled lowest Landau level. 

We focus on a simple case of such rotational symmetry breaking (used in previous theoretical investigations of gapped phases~\cite{BoYang2012}), 
introducing a Fermi surface anisotropy so that the non-interacting part of the Hamiltonian reads:
\begin{equation}
H_{0}=\frac{1}{2m_x}\Pi_x^2 + \frac{1}{2m_y}\Pi_y^2, ~~~ \Pi_i=p_i+\frac{e}{c}A_i \;,
\label{Hkin} 
\end{equation}
where $A_i$ are components of the vector potential corresponding to a uniform magnetic field $B$ along the $z$-direction. The anisotropy of the Fermi contour at zero magnetic field, $\alphaF$, is determined by the ratio of the Fermi wavevectors in perpendicular directions $x$ and $y$:
\begin{equation}
\alphaF= \frac{k_F^y}{k_F^x} = \sqrt{\frac{m_y}{m_x}}\;.
\label{alphadef}
\end{equation}
The Fermi contour achieved in the current experiments~\cite{Jo2017} is more complicated than the elliptical one represented by the above Hamiltonian; nevertheless, we expect the above model to describe the substantial $x-y$ anisotropy seen in experiment.
To the single particle Hamiltonian of Eq.~(\ref{Hkin}), we add {\it isotropic}, two-body electron-electron interactions, consistent with the appearance of fractional quantum Hall phases~\footnote{An  equivalent model may be obtained with isotropic dispersion and anisotropic interactions, as done in Ref.~\cite{Wang2012}}.

The physics of the fractional quantum Hall effect can be described through composite fermions (CFs): bound states of flux quanta and electrons~\cite{JainBook}. 
An open question is how the anisotropy of Eq.~(\ref{alphadef}) is related to the corresponding anisotropy of the composite fermions, denoted $\alphaCF$. 
Much previous work on systems described by Eq.~(\ref{Hkin}) has focussed on the Laughlin $\nu=1/3$ state. 
One can  write down model states which have a variational parameter related to the anisotropy, where Laughlin's `model wavefunction' corresponds to the isotropic case~\cite{Haldane2011,Qiu2012}. Comparing these wavefunctions to numerical exact diagonalization data allowed a determination of the relationship between $\alphaF$ and $\alphaCF$~\cite{BoYang2012}, finding an $\alphaCF$ less anisotropic than $\alphaF$.
This result is in agreement with subsequent analytical work by Murthy~\cite{Murthy2013}.
Other numerical work has found a transition out of the $\nu=1/3$ state for sufficiently large $\alphaF$~\cite{Wang2012}.

In this work, we study anisotropy at filling fraction $\nu=1/2$, where the system realizes a composite Fermi liquid phase.
By computing the anisotropy of the Fermi surface of the composite fermions, we determine the relationship between $\alphaF$ and $\alphaCF$. 
The CF Fermi surface can also be detected experimentally~\cite{Kamburov2012,Kamburov2013}, so unlike in the case for gapped states we can directly compare our results to experimental measurements, in particular experiments by the Shayegan group~\cite{Jo2017} done concurrently with this work. 
Since we obtain results for a realistic, microscopic model at $\nu=1/2$, our work is the first study of quantum Hall systems with mass anisotropy that can be directly compared to experiment.

The numerical techniques used to compute $\alphaCF$ for Laughlin states such as  $\nu=1/3$ do not apply at $\nu=1/2$,
for a number of reasons.
The variational wavefunction for a CFL has additional variational parameters representing the shape of the Fermi surface~\cite{RezayiRead}. On the finite-size systems accessible numerically, these variational parameters take discrete values, and cannot capture small changes in the anisotropy. Since the CFL is gapless, its energy spectrum is strongly dependent on size, which makes it more difficult to interpret. 
In this work we employ a different numerical technique, infinite density matrix renormalization group (iDMRG), to study a system on a cylindrical geometry of finite radius but infinite length. This technique has been successful in the study of the isotropic CFL, where a circular Fermi surface was detected~\cite{Geraedts}. 

A number of analytical results at $\nu=1/2$ are available. Ref.~\cite{Balagurov} used Chern-Simons theory to argue that $\alphaCF\propto\alphaF$, a conclusion supported by Ref.~\cite{Balram2016} where a model wavefunction satisfying this relation was proposed. Ref.~\cite{Yang2013} replaced the realistic Coulomb interaction with a Gaussian interaction, allowing an analytical calculation of  $\alphaCF$ in terms of $\alphaF$; the CFL Fermi surface was found to be less anisotropic than that of the band at $B = 0$.


\begin{figure}
\includegraphics[width=\linewidth]{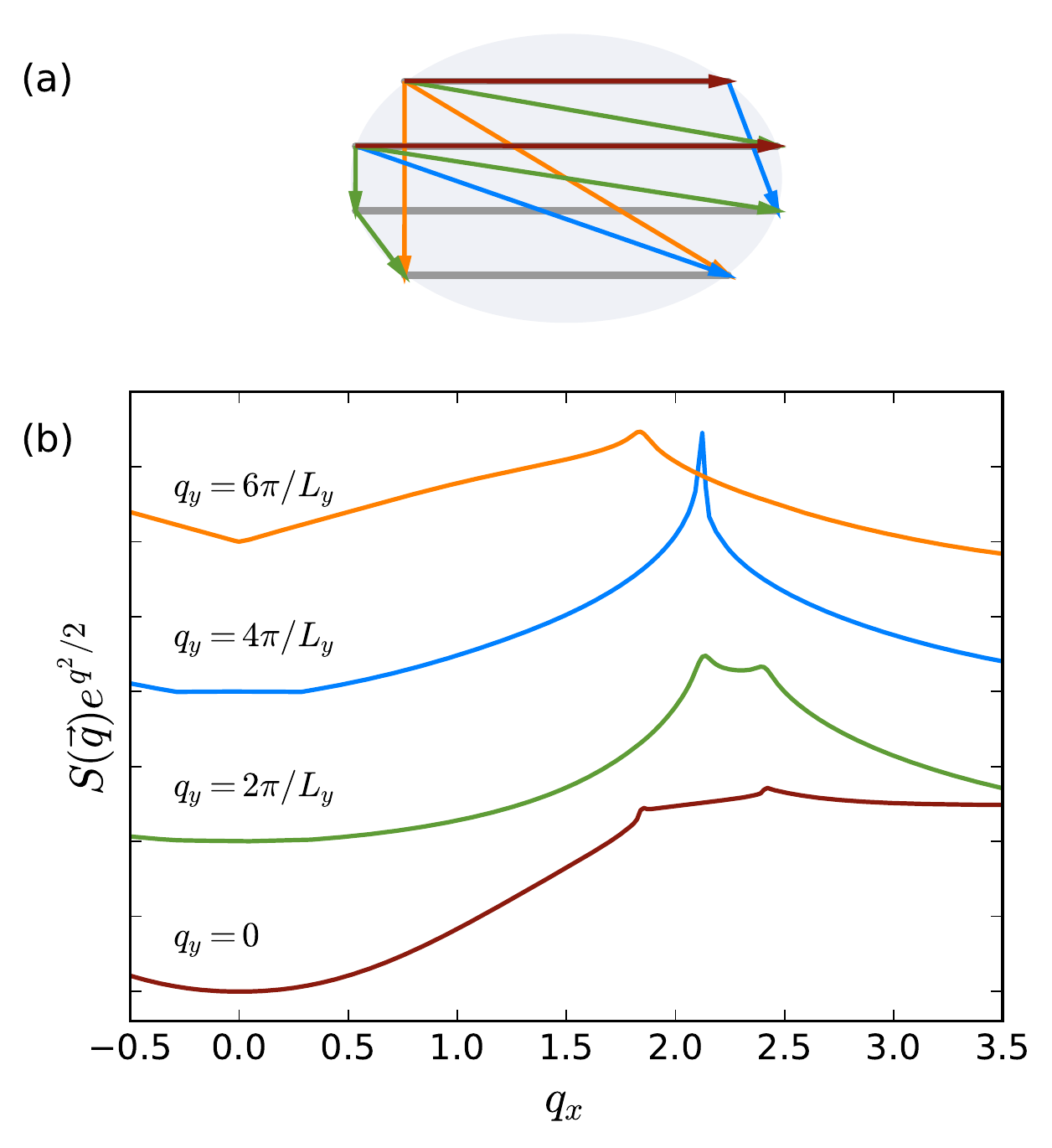}
\caption{(Color online)
Mapping the Fermi surface via the structure factor $S(\vec q)$.
(a) Fermi sea for the composite fermions at $L_y = 17 \ell_B$ and $\alphaF = 0.445$.
The shaded area represents the 2D Fermi sea in the planar ($L_y \to \infty$) limit.
The gray lines contain allowed values of $\vec q$ on the cylinder with finite $L_y$.
Colored arrows show all possible CF scattering processes between points on the Fermi surface.
(b) Numerical data for $S(\vq)$ at the lowest four allowed values of $q_y$, with bond dimension $\chi = 3000$. 
The location of the singularities can be used to determine the locations of points on the Fermi sea, and thus the CF anisotropy ($\alphaCF = 0.667$ in this case).
We multiply our data by $e^{q^2/2}$ (a smooth function of $q$) to make the singularities more clearly visible.
}
\label{method}
\end{figure}

The DMRG techniques used in this work have been described elsewhere~\cite{Zaletel2013,Geraedts}, so we provide only a brief summary here. We work on an infinite cylinder, on a spin-polarized system projected to the lowest Landau level. 
The direction along the cylinder is $x$, the one around the circumference is $y$.
DMRG is a variational technique within the ansatz of `matrix product states' (MPS), states with a limited amount of entanglement entropy.
The entanglement entropy is limited by the bond dimension, $\chi$, of the MPS. 
We project to the lowest Landau level by working in a basis containing only the lowest energy eigenstates of the Hamiltonian in Eq.~(\ref{Hkin}), which depend on the anisotropy $\alphaF$. 
We increase the bond dimension until the convergence of our wavefunctions is no longer the limiting factor on the accuracy of our results. This is achieved in most cases for $\chi$ between 3000 and 6000. 

After obtaining approximate ground states with the DMRG, we compute the guiding center structure factor:
\begin{equation}
S(\vq)\equiv\langle\rho(\vq)\rho(-\vq)\rangle,
\end{equation}
where $\rho(\vq)$ is electron density in momentum space, as a function of the wavevector $\vq$. The infinite cylinder geometry quantizes $q_y$ in steps of $2\pi/L_y$, ($L_y$ is the circumference) but allows $q_x$ to take continuous values.
Since the density of states of a Fermi liquid has a singularity at the Fermi wavevector, $S(\vq)$ will also be singular whenever the $\vq$ corresponds to a scattering process between different parts of the Fermi surface.
By computing $S(\vq)$ and locating these singularities, we can determine the shape of the Fermi surface.
Since the states obtained by DMRG are approximations of the true ground states, these singularities will not be reproduced perfectly; however, for the numerically accesible bond dimensions $\chi$, the singularities are sharp enough that the wave vectors can be identified with minimal uncertainty~\cite{Geraedts}.

Fig.~\ref{method} shows an example of how $S(\vq)$ allows us to map out the Fermi surface.
In Fig.~\ref{method}(a) the ellipse represents the anisotropic Fermi surface we are trying to investigate. The horizontal gray lines represent the allowed values of momenta in our cylinder geometry.
If we fix $q_y=0$, the horizontal arrows represent the values of $q_x$ where we expect a singularity in $S(q_x,0)$. 
We can see such singularities in the data of Fig.~\ref{method}(b). As a check, we can find singularities at other values of $q_y$, corresponding to arrows with a vertical component. 
This analysis allows us to find the set of intersection points of the gray lines in Fig.~\ref{method}(a) and the edges of the Fermi surface for any given Fermi contour anisotropy $\alphaF$ at B = 0.

We consider values of the Fermi contour anisotropy $\alphaF$ ranging from $0.16$ to $6.25$. (This corresponds to mass anisotropy range $0.025$ to $40$).
The dynamic range in $\alphaF$ is limited by convergence of the DMRG algorithm. Very small values of $\alphaF$ ($\ll 1$) increase the correlation length along the circumference of the cylinder, giving rise to increasing finite-size effects. Conversely, very large values of $\alphaF$ ($\gg 1$) increase correlations along the axis of the cylinder, requiring larger values of the bond dimension $\chi$ for convergence. This causes a rapid increase in computational time, which provides the limiting factor in that situation. While the dynamical range of $\alphaF$ is thus limited numerically, we emphasize that the range covered is significantly larger than that covered in experiment.

We consider first the experimentally relevant case of Coulomb $(1/r)$ interactions between the electrons.
In order to avoid the effects of the electron-electron interaction wrapping around the cylinder for such a long range interaction, we impose a Gaussian cutoff by multiplying our interactions by $e^{-r^2/\lambda^2}$. We take $\lambda=6\ell_B$ ($\ell_B$ is the magnetic length), a sufficiently large value that this cutoff does not affect our results.

\begin{figure}
\centering
\includegraphics[width=0.99\linewidth]{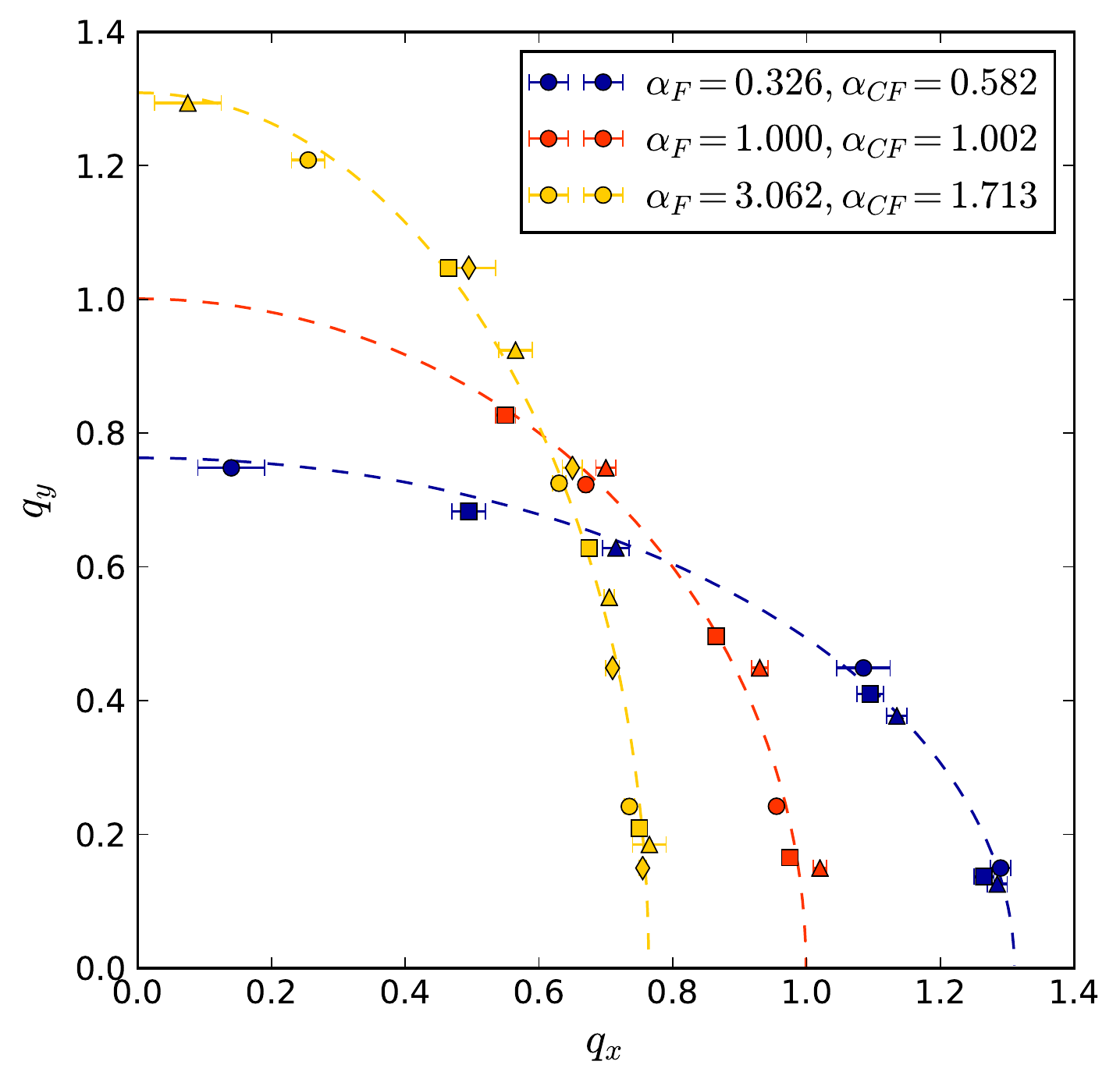}
\caption{(Color Online)
Location of the Fermi surface, for three selected values of $\alphaF$, extracted from data similar to that in Fig.~\ref{method}. 
The dashed curves are the result of fitting the data to an ellipse, with the value of $\alphaCF$ shown in the legend. Different symbols of the same color (circle, square, triangle and diamond) correspond to different system sizes but the same $\alphaF$.}
\label{ellipses}
\end{figure}

At each value of $\alphaF$, we perform the procedure described in the previous section at several values of $L_y$ (3 to 5 distinct values in the range of $13$ to $27$ magnetic lengths, depending on $\alphaF$).
This provides a list of coordinates of points which are expected to fall near the 2D Fermi contour. 
Examples representative of the cases $\alphaF <1$, $\alphaF = 1$ and $\alphaF >1$ are shown in Fig.~\ref{ellipses}.

We then extract $\alphaCF$ by fitting the resulting data to an ellipse. 
Though in the thermodynamic limit we expect an elliptical Fermi contour, our finite-size data points will deviate from this contour due to Luttinger's theorem, which in our system fixes the sum of all $q_x$ values of scattering processes with $q_y=0$ to $\nu L_y$.
This will lead to an error in the anisotropy of our fitted ellipse because of the finite size of our samples.
We can estimate this error by considering the $\alphaF = 1$ case, where $\alphaCF=1$.
Our estimate based on the total data from three system sizes is $\alphaCF = 1.002$. However, randomly removing one system size from the dataset causes the estimate to fluctuate between $0.98$ and $1.01$. 
On the basis of this result, we believe that at each value of $\alphaF$, the uncertainty on $\alphaCF$ due to Luttinger's theorem and to the finite number of system sizes considered is of order $1-2\%$.

We fit the discrete set of values of $(q_x,q_y)$ obtained this way to an ellipse of area $\pi$ from which we obtain our estimate of $\alphaCF$.
For an infinite size system in the planar limit, a $\pi/2$ rotation implies an exchange of the major and minor axes of the elliptical Fermi surface, {\it i.e.}~a change of anisotropy from $\alphaF$ to $ 1/\alphaF$, with a corresponding change of $\alphaCF$ to $ 1/ \alphaCF$. This implies that
$\alphaCF (\alphaF) \alphaCF(1/\alphaF) = 1$,
or equivalently that $\log(\alphaCF)$ is an odd function of $\log(\alphaF)$, which can be Taylor expanded~\footnote{Assuming $\alphaCF$ is analytic around the isotropic point $\alphaF = 1$.} around the isotropic point $\alphaF = \alphaCF = 1$, to yield:
\begin{equation}
\log (\alphaCF) = \gamma \log(\alphaF) + \mu \log^3 (\alphaF) + ...
\label{logeqn}
\end{equation}

\begin{figure}
\centering
\includegraphics[width=0.9\linewidth]{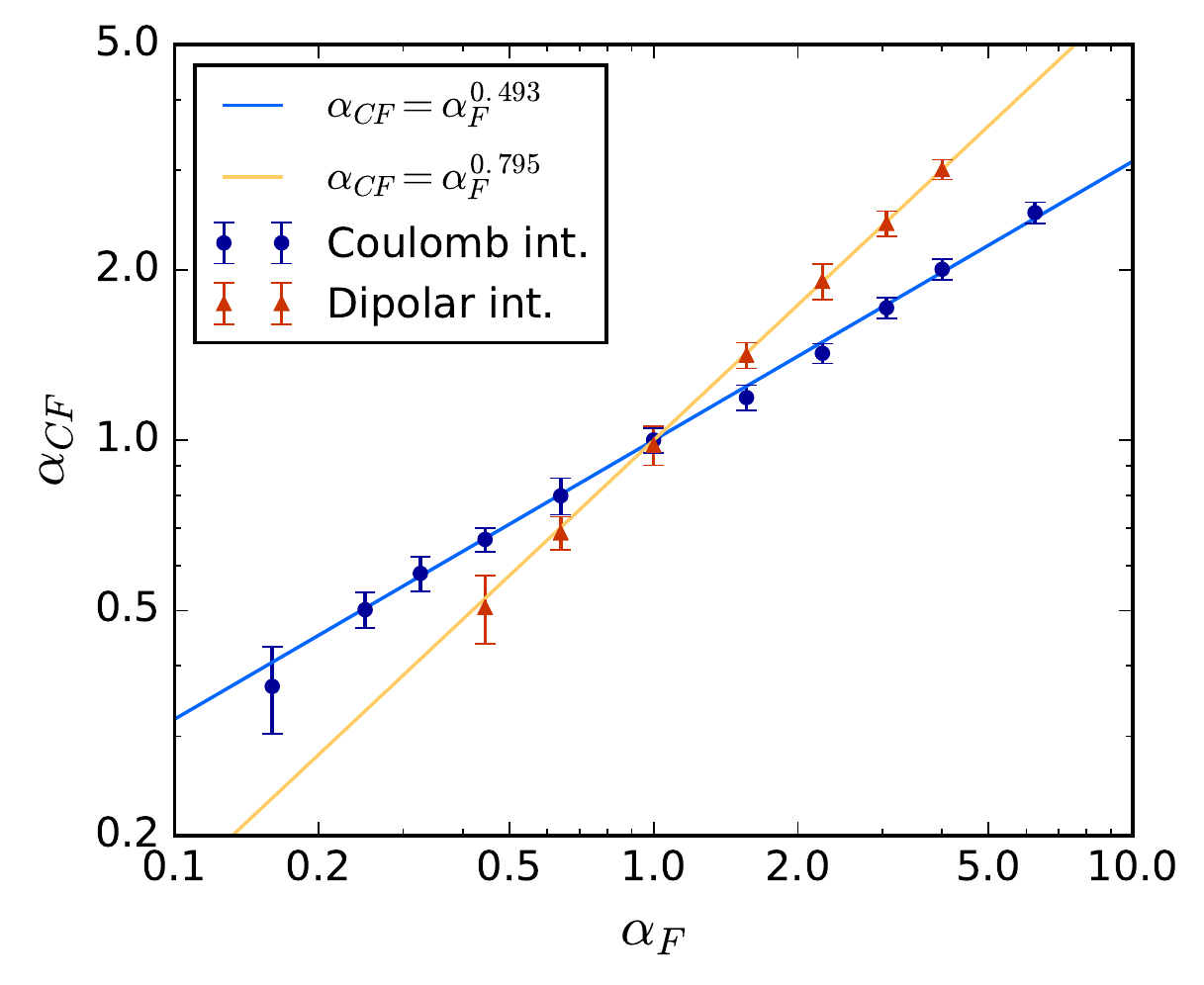}
\caption{
(Color Online)
Composite fermion anisotropy \alphaCF, computed with the method described in the main text, as a function of bare electron anisotropy $\alphaF$ for Coulomb interaction $V(r) \propto 1/r$ and dipolar interaction $V(r) \propto 1/r^3$.
The straight lines on our logarithmic axes are power-law fits $\alphaCF = \alphaF^\gamma$ with parameter $\gamma$ given in the legend.
}
\label{results}
\end{figure}

We find that the simplest such function, a power-law $\alphaCF = \alphaF^\gamma$, which corresponds to terminating the series at the first term in Eq.~(\ref{logeqn}), already fits the data well in the anisotropy range we explore (Fig.~\ref{results}).
The value of $\gamma$ we get is $\gamma = 0.493 \pm 0.008$, close to a square-root dependence $\alphaCF = \sqrt{\alphaF}$
\footnote{  An empirical fit to our data of the power-law form with adjustable exponent {\it and} prefactor, namely $\alphaCF = k {\alphaF}^\gamma$, yields $\gamma = 0.498 \pm 0.010$ and $k = 0.983 \pm 0.015$.
This form breaks the symmetry under rotation by $\pi /2$; this could be expected due to finite size and different boundary conditions (and treatments) in the $x-$ and $y-$ directions. 
Nevertheless, the exponent is very close to the one obtained imposing the symmetry, and $k$ is close to unity as required for the infinte system. }. 
Remarkably, experiments on holes in GaAs under application of in-plane strain show data in agreement with this result~\cite{Jo2017}. 
 
In order to check whether this relation is a universal feature of power-law interactions, we replace the Coulomb interaction with a dipolar interaction $V(r) \propto r^{-3}$ (also in Fig.~\ref{results}), which could be realized in a cold atomic system~\cite{Gunn1,*Gunn2,*Gunn3,Zhang2014}.
We find again that a single power law (first term in  Eq.~(\ref{logeqn}) ) fits the data well, but we measure an exponent $\gamma = 0.795 \pm 0.005 $.

This is unambiguously different from a square-root dependence, and in particular implies that $\alphaCF$ is much closer to $\alphaF$ than for the case of Coulomb interaction.
We can think of $\alphaCF$ as resulting from a competition between the non-interacting part of the Hamiltonian, with anisotropy $\alphaF$, and the interacting part of the Hamiltonian, which is isotropic.
It appears reasonable that when the Coulomb interaction is replaced with the dipolar interaction, which is weaker at long distances, the anisotropy moves closer to $\alphaF$.

\begin{figure}
\centering
\includegraphics[width=0.9\linewidth]{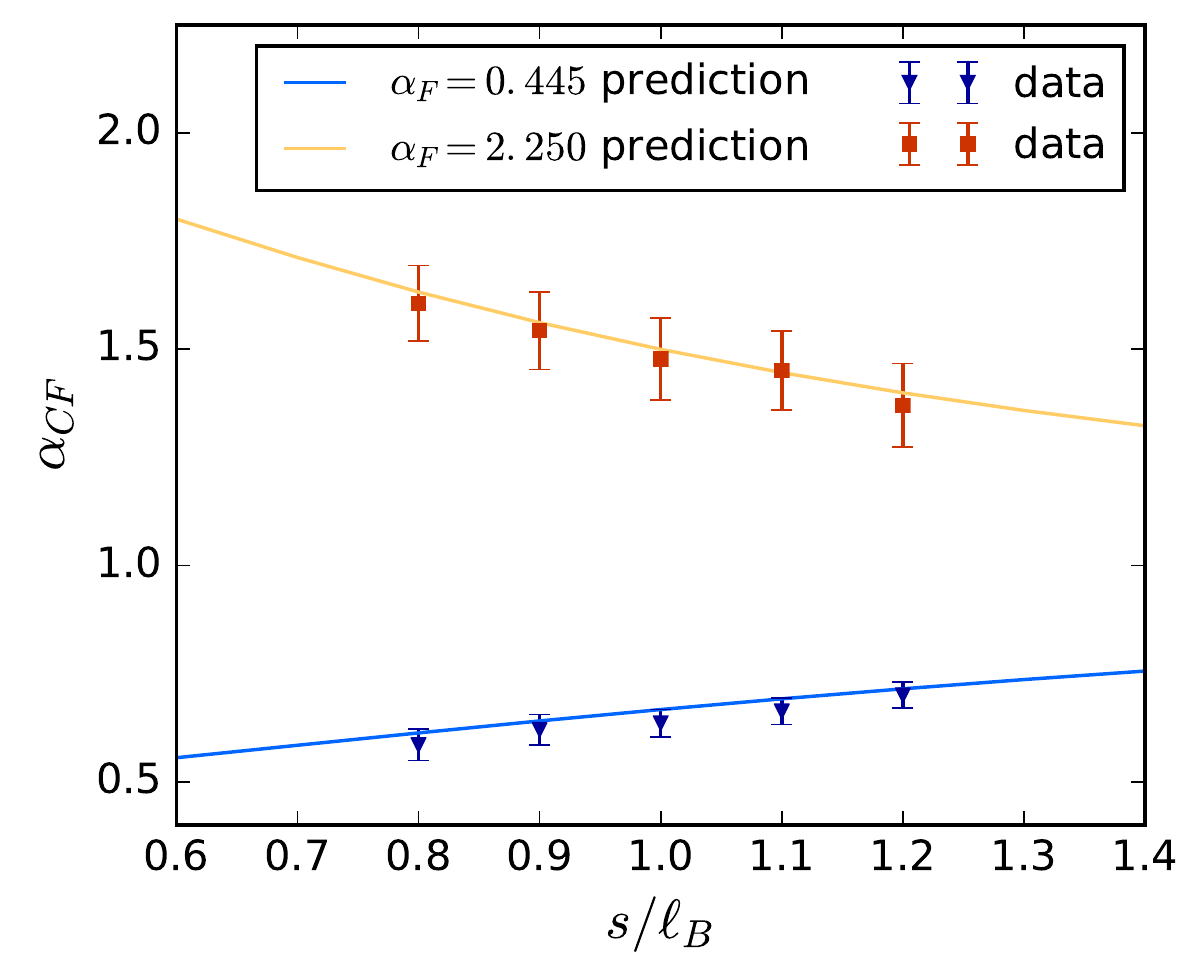}
\caption{
(Color Online)
Benchmarking our method against the exact result Eq.~\eqref{eq:yang} for a Gaussian interaction $V(r) \propto e^{-r^2/2s^2}$~\cite{Yang2013}.
We consider $\alphaF = 2.25$ and $\alphaF = 0.445$, two nearly reciprocal values to test the method in opposite regimes for varying $s$ in the vicinity of a magnetic length.
This test reveals good agreement with the analytical prediction, with a possible tendency to underestimate $\alphaCF$ by $\sim 1-2\%$.
}
\label{check}
\end{figure}

Finally, we benchmark our method against the only known exact result for $\alphaCF$: 
Yang's prediction~\cite{Yang2013} that for a Gaussian electron-electron interaction $V(r) = V_0 e^{-r^2/2s^2}$ one has
\begin{equation}
\alphaCF = \sqrt{\frac{\alphaF\ell_B^2 + s^2}{\ell_B^2 / \alphaF + s^2}} \;.
\label{eq:yang}
\end{equation}
For this purpose, we pick two nearly reciprocal values of the electron anisotropy, $\alphaF = 2.25$ and $\alphaF = 0.445$, and compute $\alphaCF$ with our method at different values of $s$ of the order of a magnetic length $\ell_B$. 
The results displayed in Fig.~\ref{check} show a good agreement with the prediction Eq.~\eqref{eq:yang}.
Our method appears to slightly underestimate $\alphaCF$ by 1 to $2\%$.
This small bias however should not significantly affect our estimates for the exponent $\gamma$ for the power-law interactions.

In summary, 
we have numerically computed the composite Fermi surface of a half-filled lowest Landau level for a two-dimensional electron gas with varying mass anisotropy, with three different forms of the electron-electron interaction. 
We find that the anisotropy of the Fermi surface of the composite Fermi liquid ($\alphaCF$) is less than that of that of the zero field Fermi surface ($\alphaF$) for non-interacting electrons. 
When the electrons in the system interact via a Coulomb interaction, our data follows the relation $\alphaCF=\sqrt{\alphaF}$. This result is in agreement with recent experimental data~\cite{Jo2017}, but not with some earlier theoretical work~\cite{Balagurov,Balram2016}. 
The relationship between $\alphaCF$ and $\alphaF$ does however depend on the form of the electron-electron interaction. For example, we find a larger composite fermion anisotropy for the $1/r^3$ interactions, and for a Gaussian interaction we find results consistent with the exact calculation in Ref.~\onlinecite{Yang2013}.

Though experiments \cite{Jo2017} find a similar relation between $\alpha_F$ and $\alpha_{CF}$ as we do, there are some differences between the two systems studied
-- experiments are conducted on quantum wells with finite width 
\footnote{This would soften the Coulomb interaction at short distances;
preliminary calculations suggest that introducing a well width of $\sim 2\ell_B$ would affect $\gamma$ by a few percent.}, 
and the experimental Fermi surface has a more complicated form than the elliptical one considered here. 
Performing simulations on a system closer to experiment is therefore an interesting extension of our research. 
Little is known about the response of the quantum Hall fluid to generalizations of Eq.~(\ref{Hkin}) ({\it e.g.} if quartic terms are added), and studies of such systems could help spur theoretical progress. 
More generally, the larger system sizes accessible in our DMRG calculations could also allow us to improve the results of previous exact diagonalization studies~\cite{BoYang2012,Qiu2012}.

{\bf Acknowledgements:}
We acknowledge helpful conversations with Insun Jo, Mansour Shayegan and Akshay Krishna.
We thank R. Mong and M. Zaletel for creating and providing the DMRG libraries used in this work; S.~D.~G. also acknowledges previous collaborations with them.  
This work was supported by Department of Energy  BES Grant DE-SC0002140.

\bibliography{anisotropy}
\end{document}